\documentclass[aps,prl,twocolumn,amsmath,superscriptaddress]{revtex4-2}
\usepackage{CJK}
\usepackage{amssymb,amsmath}
\usepackage{graphicx,color}
\usepackage{lipsum}
\usepackage{ulem}
\usepackage{hyperref}
\usepackage{ulem} 
\usepackage{enumitem}
\usepackage{lineno}

\begin{document}
\begin{CJK*}{UTF8}{gbsn}
\title{Experimental observations of marginal criticality in granular materials}

\author{Yinqiao Wang (汪银桥)}
    \affiliation{School of Physics and Astronomy, Shanghai Jiao Tong University, 800 Dong Chuan Road, 200240 Shanghai, China.}
    \affiliation{Research Center for Advanced Science and Technology, University of Tokyo, 4-6-1 Komaba, Meguro-ku, Tokyo 153-8505, Japan.}
\author{Jin Shang}
    \affiliation{School of Physics and Astronomy, Shanghai Jiao Tong University, 800 Dong Chuan Road, 200240 Shanghai, China.}
\author{Yuliang Jin}
    \affiliation{Institute of Theoretical Physics, Chinese Academy of Sciences, Beijing 100190, China.}
    \affiliation{School of Physical Sciences, University of Chinese Academy of Sciences, Beijing 100049, China.}
    \affiliation{Wenzhou Institute, University of Chinese Academy of Sciences, Wenzhou, Zhejiang 325000, China.}
\author{Jie Zhang}
    \email[Email address: ]{jiezhang2012@sjtu.edu.cn}
    \affiliation{School of Physics and Astronomy, Shanghai Jiao Tong University, 800 Dong Chuan Road, 200240 Shanghai, China.}
    \affiliation{Institute of Natural Sciences, Shanghai Jiao Tong University, 200240 Shanghai, China.}

\begin{abstract}
Two drastically different theories predict the marginal criticality of jamming. The full replica symmetry breaking (fullRSB) theory \cite{parisi2020book,parisi10review,charbonnea2014NC,charbonneau2014b} predicts the power-law distributions of weak contact forces and small inter-particle gaps in infinite-dimensional hard-sphere glass, with two nontrivial exponents $\theta_f=0.42311...$ and $\gamma=0.41269...$, respectively. While the marginal mechanical stability (MMS) analysis \cite{wyart2012,lerner2013a,degiuli2014PNAS,muller15marginal} predicts that the isostatic random packings of hard frictionless spheres under external stress are marginally stable and provides inequality relationships for the exponents of the weak-force and inter-particle-gap distributions. Here we measure precisely contact forces and particle positions in isotropic jammed bidisperse photoelastic disks and find the clear power-law distributions of weak forces and small inter-particle gaps, with both exponents $\theta_f=0.44(2)$ and $\gamma = 0.43(3) $ in an excellent agreement with the fullRSB theory. As the jammed packing subject to area-conserved cyclic pure shear approaches the yielding point, the two exponents change substantially from those of the isotropic case but they still satisfy the scaling relationship provided by the MMS argument. Our results provide strong experimental evidences for the robustness of the infinite-dimensional theory and the MMS analysis in real-world amorphous materials.
\end{abstract}

\maketitle
\end{CJK*}

\section{Introduction}
A liquid undergoes glass transition upon fast quench \cite{parisi10review,charbonnea2014NC}, whereas in the athermal situation, a flowing granular material undergoes jamming transition subjected to compression \cite{liu98diagram,ohern03jamming,liu10jamming} or shear \cite{bi11shearJamming}, with subtle structural changes while gaining rigidity. The deep connection between the two transitions has been thought for a long time \cite{liu98diagram}. Two independent theories were developed.
 
The first-principle theoretical description of amorphous materials is extremely challenging \cite{charbonneau2017review}. 
The recent fullRSB theory of the hard-sphere glass in infinite dimensions \cite{charbonnea2014NC,charbonneau2014b} unifies the glass transition and the jamming transition within the statistical mechanics framework, echoing Liu and Nagel's seminal proposal made years ago  \cite{liu98diagram}. Moreover, this theory points out a Gardner transition \cite{charbonnea2014NC,charbonneau2014b} within the glass phase similarly existing in the spin glass \cite{gardner1985}. The fullRSB theory \cite{charbonnea2014NC,charbonneau2014b} predicts that when a stable glass undergoes a Gardner transition either by compression or cooling, the metastable basins in the free-energy landscape break into a fractal hierarchy of sub-basins, forming a marginal phase. For hard spheres, the jamming transition manifested by the divergence of pressure, happens deeply inside the marginal phase. Specifically, the fullRSB solution delivers three non-trivial power-law exponents of $\kappa$, $\gamma$ and $\theta_f$ characterizing jamming, including the cage size $\Delta$ versus the pressure $p$, $\Delta\sim p^{-\kappa}$ with $\kappa = 1.41574...$, the distribution of small inter-particle gaps $P(h)\sim h^{-\gamma}$ with $\gamma = 0.41269...$, and the distribution of weak contact forces $P(f)\sim f^{\theta_f}$ with $\theta_f = 0.42311...$ \cite{charbonnea2014NC,charbonneau2014b}.

Indeed, before the fullRSB theory \cite{charbonnea2014NC,charbonneau2014b}, the relationships between these three critical exponents were already predicted by the independently developed marginal mechanical stability analysis \cite{wyart2012,lerner2013a,degiuli2014PNAS,muller15marginal}. Thus, the jamming criticality results from the remarkable convergence of two completely independent lines of research associated with two types of marginal stability \cite{franz2015,berthier2019review}. The first type refers to the (free-energy) landscape marginal stability (LMS) of the Gardner phase \cite{charbonnea2014NC,charbonneau2014b}, which resembles that of spin glass. The second type \cite{wyart2012,degiuli2014PNAS,muller15marginal} closely related to isostaticity refers to the marginal mechanical stability (MMS) of jammed packings, as introduced in the jamming field \cite{liu98diagram,ohern03jamming,wyart2005PRE,wyart2012,liu10jamming,degiuli2014PNAS}. While the low-energy excitation of Gardner phase is presumably system-wide extended in the limit of infinite dimensions \cite{charbonneau2015PRL}, both extended and localized modes may exist in low dimensions.
Besides the weak-force exponent $\theta_e$, satisfying $\theta_e = 1/\gamma-2$, which is associated with the extended excitation and coincides with the $\theta_f$ in infinite dimensions, the MMS also includes an extra exponent $\theta_l$ associated with the localized excitation, satisfying $\theta_l=1-2\gamma$ \cite{degiuli2014PNAS}. 

Since the upper critical dimension of jamming is conjectured to be two \cite{wyart2005PRE,goodrich2012}, the robustness of the fullRSB theory needs to be verified in two or three dimensions, in particular, in experiments. 
In the last decade, numerous simulations \cite{charbonneau2012,goodrich2012,lerner2012,lerner2013a,charbonnea2014NC,degiuli2014PNAS,kallus2016,charbonneau2021} have been performed. 
The exponent $\kappa$, closely related to the dynamical information of caging, has been verified in simulations of the dimension $d=3-8$ \cite{charbonnea2014NC}. Nevertheless, the two exponents $\gamma$ and $\theta_f$ can be directly obtained from a static packing.
The exponent $\gamma\approx 0.4$ of the inter-particle gaps is known to be constant in all dimensions regardless of preparation protocols \cite{charbonneau2012,donev2005,skoge2006,jin2021,lerner2013a}, except $\gamma \approx 0.5$ \cite{silbert2006,lerner2013a} when including rattlers \cite{lerner2013a}.  
However, the exponent $\theta_f$ of the weak contact forces is more sensitive to the dimension and preparation protocols, and its value has been reported within the range of 0 to 0.45 \cite{degiuli2014PNAS,ohern02,donev2005,lerner2013a,lerner2012,charbonneau2012,charbonneau2015PRL,kallus2016}. Recently, the solution of this controversy was proposed by Charbonneau et. al. \cite{charbonneau2015PRL}: if localized buckling excitation is removed the exponent of the weak force distributions coincides with the infinite-dimension solution $\theta_f = 0.42311$ in $d =2-4$.

The experimental verification of the marginal criticality of jamming is rare. Indirect characterizations of Gardner phase through particle dynamics has been performed in model glass systems, 
such as the cage separation upon compression in vibrated granular matter \cite{seguin16gardner,xiao2021}, and the logarithmic growth of the mean square displacement in colloidal glass \cite{hammond2020}. 
Yet, direct quantitatively verification of the critical exponents of the fullRSB theory \cite{charbonnea2014NC,charbonneau2014b} and the relationship between the weak force exponents and the small gap exponent \cite{wyart2012,degiuli2014PNAS,muller15marginal} in experiments is still lacking.

\section{Results}
To measure inter-particle contact forces and particle positions, we perform a two-dimensional experiment with bidisperse photoelastic disks of 1355 large disks and 2710 small disks, whose diameters are $1.4~\rm{cm}$ and $1.0~\rm{cm}$, respectively. Experimental details are described in Methods.

\textbf{Criticality of weak contact forces.}
\begin{figure*}
	\centerline{\includegraphics[width = 18 cm]{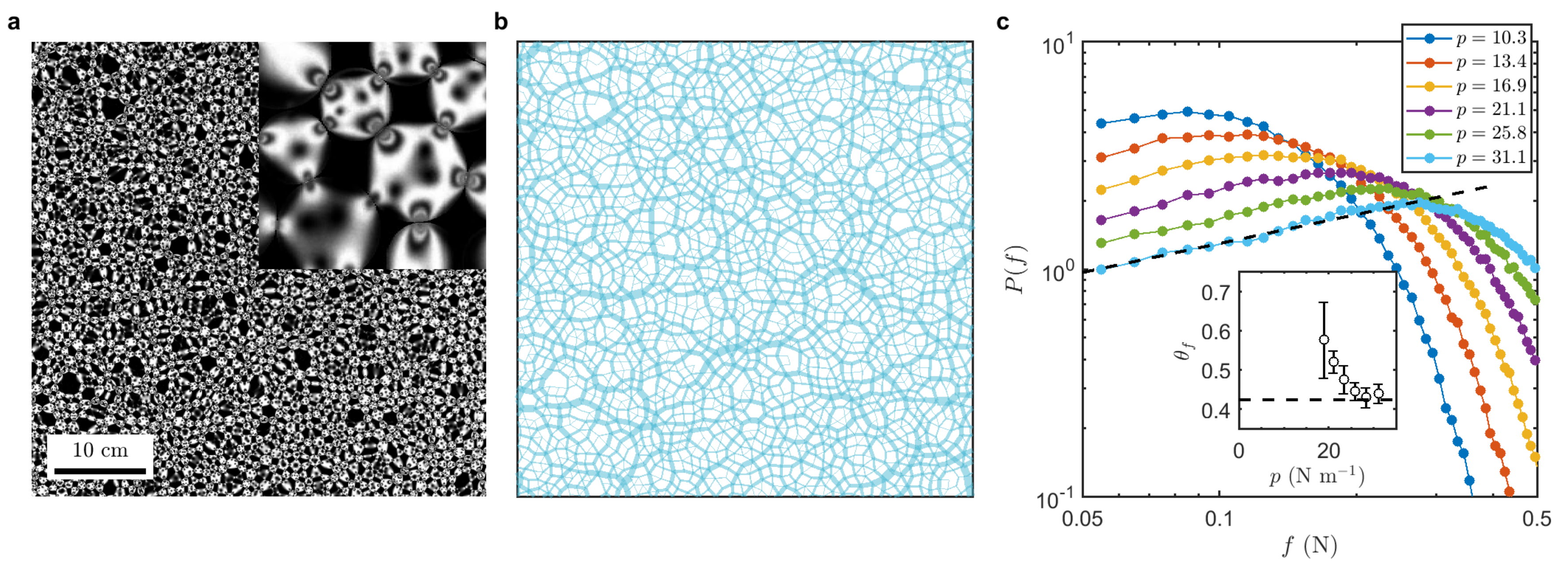}}
	\caption{\textbf{Inter-particle contact forces in compression jammed packings.} \textbf{a} A typical experimental stress image of photoelastic disks visualized by two matched circular polarizers, in which a small portion of the image is enlarged. \textbf{b} The corresponding force network is plotted, in which the thicknesses of bond are proportional to the magnitudes of the measured contact forces. \textbf{c} The probability density distributions of contact forces $P(f)$ for isotropically jammed packings at different pressure levels. The inset panel shows the fitted power-law exponent $\theta_f$ of $P(f)$ versus pressure levels, and error bars represent standard deviations of the fitted coefficients. The black dashed lines in the main panel and the inset panel represent a power law with an exponent $\theta_f = 0.42311$, which is given by the fullRSB theory.}
	\label{fig:figure1}
\end{figure*}
A typical experimental stress image is plotted in Fig.~\ref{fig:figure1}\textbf{a} and the corresponding force network is plotted in Fig.~\ref{fig:figure1}\textbf{b}. To quantify the heterogeneity of the force network, plenty of works \cite{wang2021,majmudar05nature,liu95force,tighe10review} have focused on the tail of force distribution, which is closely related to the statistical framework of granular materials.
Nevertheless, marginal criticality of jammed packings is often related to the distribution of weak contact forces, i.e. the contact forces smaller than the mean value. From the mechanical stability perspective, the weak forces are most likely to open under external perturbations to destabilize the packing. Alternatively, from the perspective of hierarchical free-energy landscape, the weak force distributions reflect the structure of the hierarchy basins in the free-energy landscape of the packing. Figure~\ref{fig:figure1}\textbf{c} shows the probability distributions of contacts forces $P(f)$ for different pressure levels on log-log scale. Here $P(f)$, as well as $g(h)$ mentioned below, are ensemble averaged over 30 jammed configurations. Limited by the measurement resolution, that is the minimum force $f \approx 0.05\ \rm{N}$ can be accurately measured, we cannot observe clear scalings of weak force distributions at low pressure. 

Subjected to isotropic compression, the global pressure of the jammed packing increases, and the range of reliable scaling becomes sufficiently large, that is $f \in [0.05, 0.6\times\langle f\rangle ]$ with $\langle f\rangle $ referring to the mean force. By fitting with a power law, the exponent $\theta_f$ at different pressure can be extracted and is plotted in the inset of Fig.~\ref{fig:figure1}\textbf{c}. We can see that the exponent $\theta_f$ saturates for large pressures, whose value shows an excellent agreement with that of the fullRSB theory \cite{charbonnea2014NC,charbonneau2014b}. The critical exponent $\theta_f = 0.42311$ is indicated as the dashed line both in the main panel and the inset of Fig.~\ref{fig:figure1}\textbf{c}. 

Note that, upon compression, even as the pressure and the contact number of the over-jammed packings exceed the values at the jamming point, the associated packings are still close to the marginal mechanical stability, which is due to the excess contact number required to stabilize the packing due to the external pressure \cite{degiuli14bp,wyart2005PRE}. In other words, a small increment of pressure tends to destabilize the packing of the fixed contact number, and hence the accompanying increase of the contact number would stabilize the packing at the new pressure level.
Further, the excess contact number reduces the fraction of bucklers significantly \cite{charbonneau2015PRL}, which explains that the measured exponent $\theta_f=0.44(2)$ in our case is close to the $\theta_e = 0.42$ but deviates substantially from the $\theta_l = 0.17$ due to the bucklers \cite{charbonneau2015PRL}.

\textbf{Criticality of small inter-particle gaps.}
\begin{figure*}
	\centerline{\includegraphics[width = 18 cm]{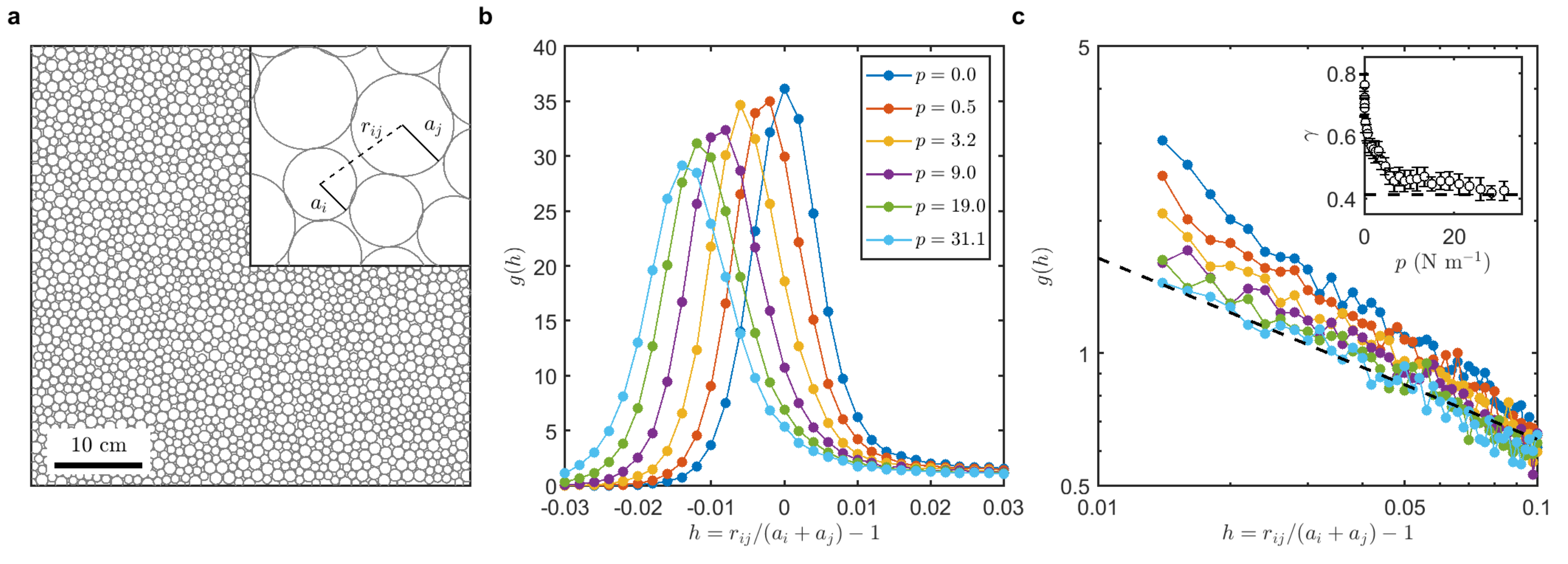}}
	\caption{\textbf{Inter-particle gaps in compression jammed packings.} \textbf{a} A snapshot of a typical bidisperse photoelastic disks packing. \textbf{b} Radial distribution functions $g(h)$ of the inter-particle gap $h$ at different pressure levels. Here $h = r_{ij}/(a_i+a_j)-1$ is the dimensionless inter-particle gap between particle $i$ and particle $j$, in which $a_i$ and $a_j$ are the radii of particles $i$ and $j$, respectively, and $r_{ij}$ is the distance between particles $i$ and $j$. \textbf{c} $g(h)$ plotted on a log-log scale. The inset panel shows the power-law exponent $\gamma$ of $g(h)$ fitted in the range [0.012,0.05] versus the pressure levels. The black dashed lines in the main panel and the inset panel represent a power law with an exponent $\gamma = 0.41269$, which is given by the fullRSB theory.}
	\label{fig:figure2}
\end{figure*}
While the opening of contacts bearing weak forces destabilizes the packing, the nearly-touched contacts can close the gaps to form new contacts to stabilize the packing. For mono-disperse packings, the distribution of small gaps can be directly given by the radial distribution function $g(r)$.
Due to the bidisperse characteristic in our experiments, the first peak of $g(r)$ will split into three peaks. To avoid the effects of bidispersity, we define the dimensionless inter-particle gaps as $h=r_{ij}/(a_i+a_j)-1$, here $r_{ij}$ is the distance between two neighbouring particles $i$ and $j$, and $a_i$ and $a_j$ are the radii of the two particles, respectively, as shown in Fig.~\ref{fig:figure2}\textbf{a}. Finally, we obtain the distributions of dimensionless inter-particle gaps $g(h)$, as shown in Fig.~\ref{fig:figure2}\textbf{b} and \textbf{c}.

For sphere packings near the jamming transition, $g(h)$ is a delta function at $h=0$ followed by a power-law decay $g(h)\propto h^{-\gamma}$ \cite{donev2005,silbert2006}. 
However, the delta function often smears out due to the uncertainty of particle positions in experiments, as shown in Fig.~\ref{fig:figure2}\textbf{b}. The width of the Gaussian-like peak at $h=0$ is around 0.01, which indicates the uncertainty of the particle detection is around half pixel. 
Even if the small inter-particle gaps had a power-law scaling, it would be hidden by the Gaussian-like peak near $h=0$. For soft sphere packings, the peak of $g(h)$ shifts down upon compression, as shown in Fig.~\ref{fig:figure2}\textbf{b}. Therefore, applying compression on the packings helps to disentangle the Gaussian-like peak and the scaling of positive inter-particle gaps with $h>0$. In Fig.~\ref{fig:figure2}\textbf{c}, we plot $g(h)$ on log-log scale and find an increasingly clear power-law scaling as pressure increases. As the very small gaps are interfered by the accuracy of particle positions and meanwhile the critical scaling may break down for large gaps, we fit the $g(h)$ in the range $h\in [0.012,0.05]$ to the power law. The fitted exponents are shown in the inset of Fig.~\ref{fig:figure2}\textbf{c}. 
Similar to the exponent $\theta_f$ of weak forces, the exponent $\gamma$ is also in good agreement with the fullRSB solution, indicated by the dashed lines in Fig.~\ref{fig:figure2}\textbf{c}. Note that the fraction of rattlers is around 0.8\% in the over-jammed packings, thus negligible.

\textbf{Non-universal criticality in sheared packings with friction.}
\begin{figure*}
	\centerline{\includegraphics[width = 18 cm]{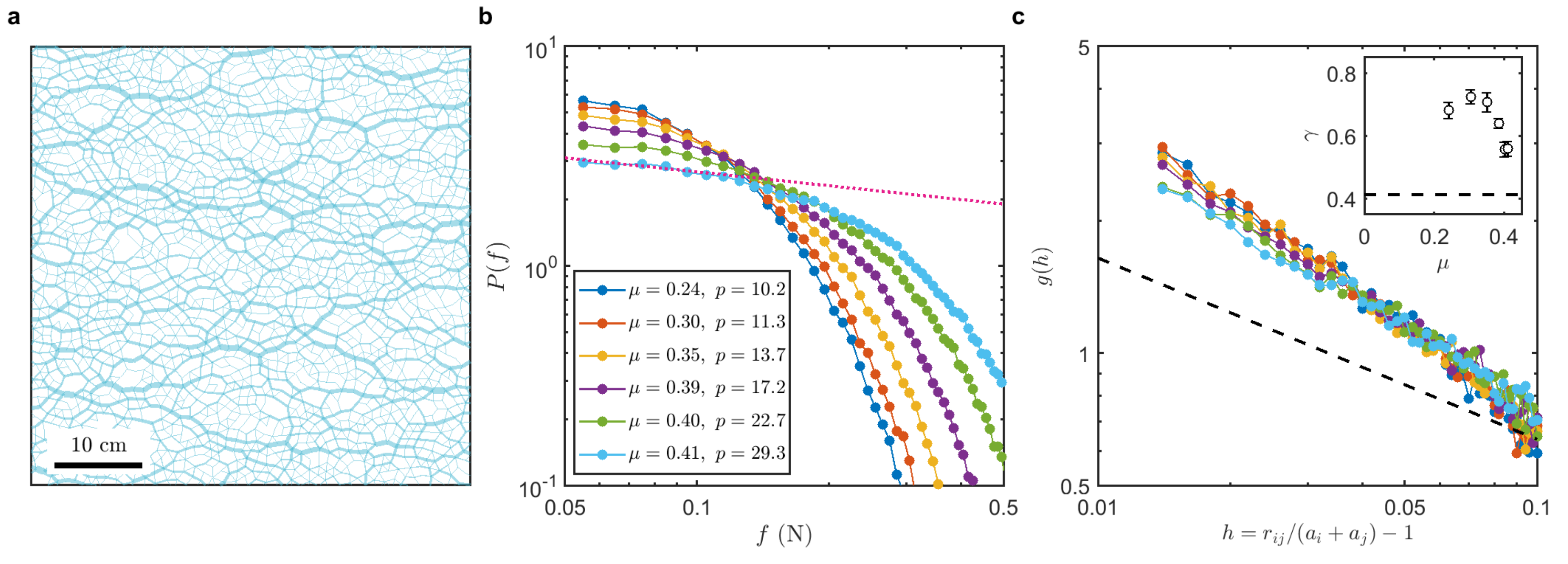}}
	\caption{\textbf{Inter-particle forces and gaps in jammed packings subject to cyclic shear.} \textbf{a} A typical force network for jammed packings subject to cyclic pure shear. \textbf{b} Distributions of inter-particle forces $P(f)$, and \textbf{c} Distributions of inter-particle gaps $g(h)$ for jammed packings with different global stress ratios $\mu$. The red dotted line represents the power law with an exponent $\theta_f=-0.21$, which is derived from the exponent $\gamma=0.56$ and the scaling relationship given by the marginal mechanical stability analysis. The black dashed lines in the panel \textbf{c} represents a power law with an exponent $\gamma = 0.41269$, as a guide for the eye.}
	\label{fig:figure3}
\end{figure*}
In simulations of hard-sphere glass, the packing fraction of jamming point $\phi_J$ depends on the parent liquid state and the preparation protocol, which results in the jamming line \cite{charbonnea2014NC,jin2021}. Specifically, a denser parent liquid produces a denser jammed packing \cite{charbonnea2014NC}, and applying shear further expands the phase space of jammed states \cite{jin2021}. Nevertheless, the marginal critical scalings associated with jamming are universal regardless of the path to jam, that the associated exponents, including $\gamma\approx0.4$ \cite{jin2021,babu2022}, $\theta_e\approx0.42$ and $\theta_l\approx0.17$ \cite{babu2022}, remain unchanged. Next we will show however that for jammed frictional particles the situation becomes different.

Figure \ref{fig:figure3}\textbf{a} shows a typical force network of the jammed packings prepared under the steady-state cyclic pure shear, in which the global stress ratio $\mu$ (See Methods for details) is around 0.41. Both the weak force distributions $P(f)$ and the small inter-particle gap distributions $g(h)$ depart obviously from the infinite-dimensional solution, as shown in Figs.~\ref{fig:figure3}(b-c). This might explain the relatively large range of the exponent $\gamma\in [0.25,0.75]$ reported in colloidal \cite{kyeyune2018} and granular experiments \cite{aste2005} under gravity. 
While $P(f)$ typically shows a peak around the mean contact force for isotropically jammed packings as shown in Fig.~\ref{fig:figure1} and also in literature \cite{ohern02,wang2021}, the peaks disappear here in Fig.~\ref{fig:figure3}: for jammed packings subject to the steady-state cyclic pure shear, within our contact-foce resolution,  $P(f)$ increase monotonically as $f$ decreases.

Remarkably, even both the weak forces exponent $\theta_f$ and the small gaps exponent $\gamma$ deviate substantially from those of the infinite-dimensional solution, they still satisfy the relationship predicted by the MMS analysis \cite{wyart2012,lerner2013a,degiuli2014PNAS,muller15marginal}. 
Here, the global stress ratio $\mu=0.41$ of the jammed packings subject to the steady-state cyclic pure shear is close to the yielding point. Figure \ref{fig:figure3}\textbf{c} shows that the exponent $\gamma$ of small gaps is around 0.56. 
According to the theory \cite{wyart2012,degiuli2014PNAS,muller15marginal}, the exponent $\gamma=0.56$ derives the two weak force exponents $\theta_l = 1-2\gamma=-0.12$ and $\theta_e= 2-1/\gamma=-0.21$, which then determine the weak-force exponent $\theta_f$ according to the relationship $\theta_f = \min(\theta_l,\theta_e) = \theta_e = -0.21$.
As shown in Fig.~\ref{fig:figure3}\textbf{b}, the $P(f)$ of the packing $\mu=0.41$ is very close to the power-law scaling with an exponent $\theta_f=-0.21$, plotted as the red dotted line, which indicates that the bound between $\theta_f$ and $\gamma$ is marginally satisfied.
Moreover, as estimated from $\gamma$, the fact that $\theta_e < \theta_l$ suggests that the extended excitation becomes dominant in the weak force region for jammed packings subject to the steady-state cyclic pure shear near yielding.
Note that the systematic evaluation of $\theta_f$ deteriorates rapidly for $\mu<0.41$ since the pressure drops with $\mu$ and hence the power-law regime narrows substantially.

\section{Discussion}
We have experimentally investigated  the marginal criticality in jammed bidisperse photoelastic disks.

First, we have measured the power-law exponents of the weak-force distributions and the small inter-particle gap distributions in the isotropic jammed packings. As the global pressure increases, the interval of the weak forces (i.e. $f<\langle f \rangle$) above the minimum force resolution increases significantly, and meanwhile the localized excitation is suppressed. We thus can obtain a reliable measurement of the weak force exponent associated with extended excitation, i.e. $\theta_f = \theta_e= 0.44(2)$, and the small inter-particle gap exponent $\gamma=0.43(3)$, both of which agree extremely well with those of the fullRSB solution \cite{charbonnea2014NC,charbonneau2014b}.

Next, we investigate the two exponents in the jammed packings produced in the steady-state cyclic pure shear. We find that both exponents of $\theta_f$ and $\gamma$ deviate substantially from those of the isotropic case, especially when the stress ratio $\mu$ approaches the yielding point, the accompanied pressure value increases significantly due to shear-induced dilation that creates reasonably wide ranges of power-law scaling. Remarkably, the two exponents still satisfy the relationship predicted by the marginal mechanical stability \cite{wyart2012,degiuli2014PNAS,muller15marginal}. Furthermore, the extended modes dominate over the low-energy excitation based on the estimation from $\gamma$, implying that the avalanche dynamics of granular materials shall percolate the whole system near the yielding point.

Note that the inter-particle friction in our system is beyond the present scope of the fullRSB theory \cite{charbonnea2014NC,charbonneau2014b}. We believe that friction expands the phase space of the jammed states significantly compared with that of the frictionless spheres or disks. Recall that to create the isotropic jammed packings, we first prepare an initial stress-free packing with the packing fraction $\phi=83.4\%$, close to the two-dimensional jamming point of frictionless particles $\phi_J\approx 84\%$ \cite{ohern03jamming}, and then we apply constant vibrations at the bottom plate to eliminate the base friction. Adopting this preparation protocol allows us to generate the almost perfectly uniform and isotropic jammed packings \cite{wang2020shearlocalization,bulbul20elasticity}, from which we obtain the good statistics of the contact forces and inter-particle gaps that are consistent with the prediction of frictionless hard spheres \cite{charbonnea2014NC,charbonneau2014b}. To create the jammed packings upon the steady-state pure shear of conserved area, we believe that the inter-particle friction plays an important role in selecting the jammed states in the expanded phase space, which therefore changes the exponents of $P(f)$ and $P(h)$.

It has been reported that the Gardner phase is not necessarily universally observed \cite{scalliet2017noG,albert2020} in finite-dimensional glasses far away from jamming. The fact that the jamming criticality are observed in low dimensions might be due to the emergence of hyperuniformity \cite{rissone2021,wilken2020hyperuniform,wilken2021,torquato11hyperuniform} in jammed packings, which may suppress the fluctuations in finite-dimensional systems. 
Meanwhile, a clear theoretical basis for the super-universality of jamming is still lacking \cite{goodrich2012,berthier2019review,liu10jamming,hexner2019}.

In the future, the connection between these critical scalings of micro structures and the rheological dynamics of amorphous solids need to be studied. In addition, the effects of particle shape on the micro structures of inter-particle gaps and contact forces \cite{brito18} can be explored using the photoelastic ellipses \cite{wang2021}.

\textbf{Acknowledgments} Y.W., J.S. and J.Z. acknowledge the support from the NSFC (No. 11974238 and No. 11774221) and from the Innovation Program of Shanghai Municipal Education Commission under No. 2021-01-07-00-02-E00138. Y.W., J.S. and J.Z. also acknowledge the support from the Student Innovation Center of Shanghai Jiao Tong University. Y.W. acknowledges the support from Shanghai Jiao Tong University via the scholarship for outstanding Ph.D. graduates.
Y.J. acknowledges the support from the NSFC(No. 11974361, No. 12161141007, No. 11935002 and No. 12047503), from the Key Research Program of Frontier Sciences, Chinese Academy of Sciences, Grant NO. ZDBS-LY-7017, and from the Key Research Program of the Chinese Academy of Sciences, Grant NO. XDPB15.
	
\textbf{Competing Interests} The authors declare that they have no competing financial interests.
	
\textbf{Author Contributions} Y.W. and J.Z. conceived the project and designed the experiment. Y.W. and J.S. performed the experiments and analysed the data. Y.W., J.S., Y.J. and J.Z. contributed to writing the paper.

\textbf{Correspondence and requests for materials} should be addressed to J.Z. (email: jiezhang2012@sjtu.edu.cn).

\normalem
%

\appendix
\section{Methods}

The layer of particles is placed on a horizontal glass plate, and is laterally confined by two pairs of walls, forming a rectangle. The two pairs of walls can move inward simultaneously to apply isotropic compression, or one pair moves inward and the other pair moves outward to apply area-conserved pure shear. To eliminate the friction between the particle layer and the glass plate, we attach eight mini vibrators below the glass plate.
At the top, an array of 2$\times$2 high-resolution (100 pixel/cm) cameras are aligned and synchronized.
A pair of matched circular polarizers is inserted above and below the particle layer. The top polarizer is right below the array of cameras and can be moved horizontally in and out under the control of a motor. A uniform green light box is mounted at the bottom.

Packings at different pressure levels subject to isotropic compression are prepared as follows. We first prepare a stress-free random and homogeneous particle configuration with the packing fraction $\phi=83.4\%$ (the ratio between the area of disks and that of the rectangle) near the jamming point $\phi_J\approx 84.0\%$ of frictionless particles \cite{ohern03jamming}. To achieve this goal, we start from a random loose initial configuration by filling a random mixture of bidisperse photoelastic disks within the rectangle. When applying isotropic compression to increase the $\phi$, we constantly apply manual agitation within the photoelastic disks to remove any transient force chains and meanwhile turn on the mini vibrators to eliminate the base friction.
After the stress-free random initial packing with $\phi=83.4\%$ is prepared, we stop manual agitation for any further isotropic compression. 
Next, we apply isotropic compression in incremental steps to obtain packings at a sequence of incremental pressure levels with maximum packing fraction $\phi=86.2\%$. During this process, the mini vibrators are synchronized with the movement of the four walls such that they are turned on when the walls move and are turned off when the walls pause after an incremental step. The force chains of isotropically jammed packings are homogeneous and isotropic. An example of an original stress image and the associated force-network image are shown in Fig.~\ref{fig:figure1}(a-b). Note that we can generate an ensemble of sets of isotropically jammed packings by applying the above protocol repetitively. 

Packings of different pressure levels subject to area-conserved pure shear are prepared as follows. We first follow the protocol of preparing a weakly jammed packing subject to isotropic compression with the packing fraction $\phi=85.0\%$. We then apply pure shear to this packing to drive the system into steady states, in which the curve of the global stress ratio $\mu\equiv\tau/p$ versus strain of every shear cycle remains fixed. Here the global stress ratio $\mu$ refers to the ratio between the global shear stress $\tau$ and the pressure $p$ of the system. The strain $\epsilon$ of every shear cycle is in between $\epsilon_{min}=-2.8\%$ and $\epsilon_{max}=2.8\%$. We apply pure shear in incremental steps to cover a sequence of strains within  $\epsilon_{min}\le\epsilon\le\epsilon_{max}$. During this process, the mini vibrators are synchronized with the movement of the four walls such that they are turned on when the walls move and are turned off when the walls pause after an incremental step. Since there is shear induced dilation in area-conserved cyclic pure shear, the pressure $p$ of the system is not a constant, and hence we obtain a sequence packings of different $\mu$ and $p$ subject to cyclic pure shear. Note that we can similarly generate an ensemble of sets of cyclically sheared jammed packings by applying the above protocol repetitively. 

For each step of compression or shear, we record one stress image, as shown in Fig.~\ref{fig:figure1}\textbf{a}, and one normal image depending on whether the top polarizer is inserted or removed. With the normal image, we use the Hough transform algorithm to detect the positions of particles with sub-pixel resolutions. We can then obtain inter-particle contacts with the criterion of $r_{ij} < (1+\delta)(a_i+a_j)$. Here $r_{ij}$ is the distance between particle $i$ and particle $j$, $a_i$ and $a_j$ are the radii of particles, and $\delta$ is set as 0.05 to avoid the missing of force-bearing contacts. Note that although this value of $\delta$ may include a few false contacts, it however will not affect the results presented in this paper since the measured contact forces of those false contacts are smaller than 0.05 N. 
With the stress image and the positions of particles and contacts, we can solve the contact force vectors using a force-inverse algorithm, which generates a computed stress image based on an initial guess of contact forces, and then iterate to minimize the difference between the experimental and computed stress images.
The relative error of contact force measurement is less than 5\% for the typical force magnitude; When $f < 0.05\ \rm{N}$, measurements are often interfered by the inhomogeneity of background illuminations and limited by the image resolution. More details of the experimental setup and the force-inverse algorithm can be found in ref. \cite{wang2021}. Our final remark is regarding the mini vibrators: Applying vibration using mini vibrators efficiently eliminates the base friction but this method will not produce noticeable effects in the relaxation of the stress images. For a jammed packing, we turn on the mini vibrators for a considerably long time and we compare the stress images before and after see no noticeable changes. 

\end{document}